\documentclass[twocolumn,showpacs,amsmath,amssymb]{revtex4}
\usepackage{graphicx}% Include figure files
\usepackage{dcolumn}% Align table columns on decimal point
\usepackage{bm}% bold math

\begin{document}
\title{A model for cascading failures in complex networks} 

\author{ Paolo Crucitti$^{1}$, 
         Vito Latora$^{2}$ and 
         Massimo Marchiori$^{3}$  }

\affiliation{\vspace{.3 cm}$^{1}$Scuola Superiore di Catania,
Via S. Paolo 73, 95123 Catania, Italy}
\affiliation{$^{2}$ Dipartimento di Fisica e Astronomia, 
Universit\`a di Catania, and INFN sezione di Catania, 
Via S. Sofia 64, 95123 Catania, Italy}
\affiliation{$^{3}$W3C and Lab. for Computer Science,
Massachusetts Institute of Technology, USA}

\begin{abstract}
Large but rare cascades triggered by small initial shocks 
are present in most of the infrastructure networks. 
Here we present a simple model for cascading failures 
based on the dynamical redistribution of the flow on the network.  
We show that the breakdown of a single node is sufficient to 
collapse the efficiency of the entire system if the node is among  
the ones with largest load. 
This is particularly important for real-world networks with a highly 
hetereogeneous distribution of loads as the Internet 
and electrical power grids. 
\end{abstract}
\maketitle

Cascading failures are common in most of the complex 
communication/transportation networks \cite{mendesbook,strogatzreview} 
that are the basic components of our lives and industry. 
In fact, although most failures emerge and dissolve locally, 
largely unnoticed by the rest of the world, a few trigger 
avalanche mechanisms that can have large effects over 
the entire networks. 
\\
Cascading failures take place on the Internet, 
where traffic is rerouted to bypass malfunctioning routers, 
eventually leading to an avalanche of overloads on other routers 
that are not equipped to handle extra traffic. 
The redistribution of the traffic can result in a congestion 
regime with a large drop of the performance. 
For instance in October 1986, during the first documented 
Internet congestion collapse, 
the speed of the connection between the Lawrence Berkeley 
Laboratory and the University of California at Berkeley, 
two places separated only by 200 meters, dropped by 
a factor 100 \cite{jacobson,guimera}.  
\\
Cascading failures also take place in electrical power grids. 
In fact, when for any reason a line goes down, its power is 
automatically shifted to the neighboring lines, which in most 
of the cases are able to handle the extra load. 
Few times, however, also these lines are overloaded  
and must redistribute their increased load to their neighbors. 
This eventually leads to a cascade of failures: a large 
number of transmission lines are overloaded and malfunction 
at the same time.  
This is exactly what happened in August 10 
1996 \cite{carreras,sachtjen} when 
a 1,300-megawatt electrical line in southern Oregon 
sagged in the summer heat, initiating a chain reaction that cut 
power to more than 4 million people in 11 Western States. 
And probably this is also what happened in August 14 2003 
when an initial disturbance in 
Ohio \cite{NYT} triggered the largest blackout in the US's history 
in which millions of people remained without 
electricity for as long as 15 hours. 
\\
Large cascading failures are also present in social and 
economic systems \cite{watts_failures}.   

How is it possible that a small initial shock,  
as the breakdown of an Internet router 
(or of an electrical substation or line), can trigger 
avalanches mechanisms affecting a considerable fraction 
of the network and collapsing a system 
that in the past was proven to be stable 
with respect to similar shocks ? 
In this letter we propose a simple model for 
cascading failures in complex networks. 
Resistance of networks to the removal of nodes 
or arcs, due either to random breakdowns or to intentional 
attacks has been studied in 
Refs.\cite{barabasi_attack,holme,crucitti,girvan,motter2002}. 
Such studies have focussed only on the {\it static properties} of 
the network showing that the removal of 
a group of nodes altogether can have important consequences.   
%
%is robust against random failure of nodes 
%and to some extent is even robust against intentional attacks.  
%
Here we show how the breakdown of a {\it single node} 
is sufficient to collapse 
the entire system simply because of the 
{\it dynamics of redistribution of flows} on the network.  
In our model each node is characterized by a given
{\it capacity} to handle the traffic. Initially 
the network is in a stationary state in which 
the {\it load} at each node is smaller 
than its capacity. The breakdown (removal) of a node 
changes the balance of flows and leads to a  
redistribution of loads over other nodes. 
If the capacity of these nodes can not handle 
the extra load this will be redistributed in turn,   
triggering a cascade of overload failures 
and eventually a large drop of the network 
performance as the ones observed in real systems, like the Internet or 
the electrical power grids.  
The main differences with respect to previous 
models \cite{motter_cascade,moreno_vespignani,fiberbundle} are: 
\\
1) overloaded nodes are not removed from the network. 
It is the communication passing through overloaded (congested) 
nodes that will get worse, so that eventually  
the information/energy will avoid congested nodes.  
\\
2) the damage caused by a cascade is quantified in terms 
of the decrease in the network {\it efficiency}, a variable 
defined in Ref.\cite{lm2}.  
\\
First we introduce the model and then we show some applications 
to artificially created topologies, to the Internet and to 
the electrical power grid of the western United States.

\bigskip
We represent a generic communication/transportation network as a 
valued (weighted) \cite{wasserman} undirected \cite{note0} graph $\bf G$,  
with $N$ nodes (the internet routers or the substations of 
an electrical power grid) and $K$ arcs (the transmission 
lines). 
$\bf G$ is described by the $N \times N$ adjacency matrix 
$\{ e_{ij} \}$. If there is an arc between node $i$ and node $j$, 
the entry $e_{ij}$ is the value, a number in the range (0,1] 
attached to the arc, otherwise $e_{ij} =0$ \cite{note1}. 
Such a number is a measure of the efficiency in the communication 
along the arc. For instance, in the Internet, the smaller   
$e_{ij}$ is, the longer it takes to exchange an unitary packet of 
information along the arc between $i$ and $j$. 
Initially, at time $t=0$, we set $e_{ij} = 1$ for 
all the existing arcs, meaning that all the transmission lines 
work perfectly and are equivalent. 
The model we will propose consists in a rule for the time evolution 
of $\{ e_{ij} \}$ that mimics the dynamics of flow redistribution 
following the breakdown of a node. 
To define the network efficiency \cite{lm2} we assume that 
the communication between a generic couple of nodes 
takes the most efficient path connecting them. 
The efficiency of a path is the so-called 
harmonic composition \cite{performance1, performance2,hc} 
of the efficiencies of the component arcs. 
By $\epsilon_{ij}$ we indicate the efficiency of the most efficient 
path between $i$ and $j$. Matrix $\{\epsilon_{ij}\}$ is calculated 
by means of the algorithms used in Ref.\cite{lm2}. 
Then the average efficiency of the network is: 
\begin{equation}
\label{efficiency}
E({\bf G}) =           \frac{1}{N(N-1)}
{\sum_{{i \ne j\in {\bf G}}} \epsilon_{ij}}
\end{equation}
and is used as a measure of the performance of ${\bf G}$ at a given time. 
\\ 
The {\it load} $L_i(t)$ on node $i$ at time $t$ is the total number 
of most efficient paths passing through $i$ at time $t$ \cite{goh}. 
Each node is characterized by a {\it capacity} defined  
as the maximum load that node can handle.
Following Ref.\cite{motter_cascade} 
we assume the capacity $C_i$ of node $i$ 
proportional to its initial load $L_i(0)$: 
\begin{equation}
C_i = \alpha ~\cdot L_i(0) ~~~~ i=1,2,...N 
\label{capacity}
\end{equation}
where $\alpha \ge 1$ is the tolerance  
parameter of the network \cite{note2}. 
This is a realistic assumption in the design of an 
infrastructure network, since the capacity can not 
be infinitely large because it is limited by the cost. 
With such a definition of capacity, the network we 
have created is in a stationary state in which it operates with a 
certain efficiency $E$.   
The initial removal of a node \cite{note3}, simulating 
the breakdown of an Internet router or of an electrical substation, 
starts the dynamics of redistribution of flows 
on the network. In fact the removal of a node 
changes the most efficient paths between nodes and consequentely 
the distribution of the loads, creating overloads  
on some nodes. 
At each time $t$ we adopt the following iterative rule: 
\begin{equation}
e_{ij}(t+1)  =\left\{ \begin{array}{cc}
	e_{ij}(0) \cdot \frac {C_i} {L_i(t)}    & ~~~$if$~~   L_i(t) > C_i \\
	e_{ij}(0)    &~~~ $if$~~  L_i(t) \le C_i
       \end{array}
\right. 
\label{dynamics}
\end{equation} 
where $j$ extends to all the first neighbours of $i$. 
In this way if at time $t$ a node $i$ is congested,  
we reduce the efficiency of all the arcs  passing through it, 
so that eventually the information/energy will take alternative 
paths (the new most efficient paths). 
This is a softer and, for some applications, 
a more realistic situation than the one considered 
in Ref.\cite{motter_cascade}, in which the 
overloaded nodes are removed from the network. 
Rule (\ref{dynamics}) produces a decrease of the efficiency of 
the network $E$ and, as we will show in the following, 
in some cases it can trigger an avalanche mechanism 
collapsing the whole system.

\begin{figure}[!h]
\includegraphics[width=5.0cm,angle=0]{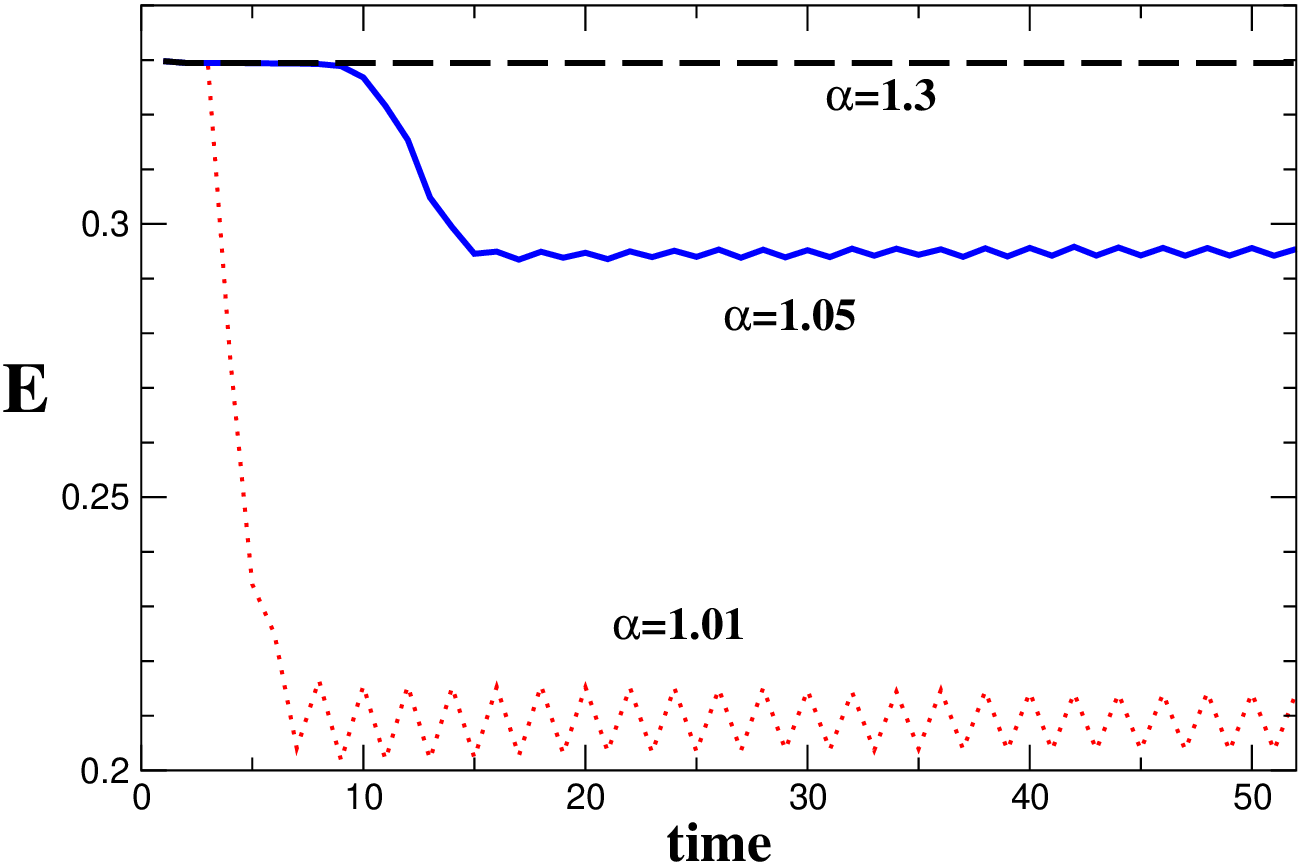}   
\caption{\label{fig1} 
Cascading failure in a BA scale-free network as triggered by the 
initial removal of a single node chosen at random.  
We plot the efficiency $E$ of the network as a function of the 
time for three values of the tolerance parameter $\alpha$. 
The curves correspond to an average over ten triggers.}
\end{figure}

\begin{figure}[!h]
\includegraphics[width=6.5cm,angle=0]{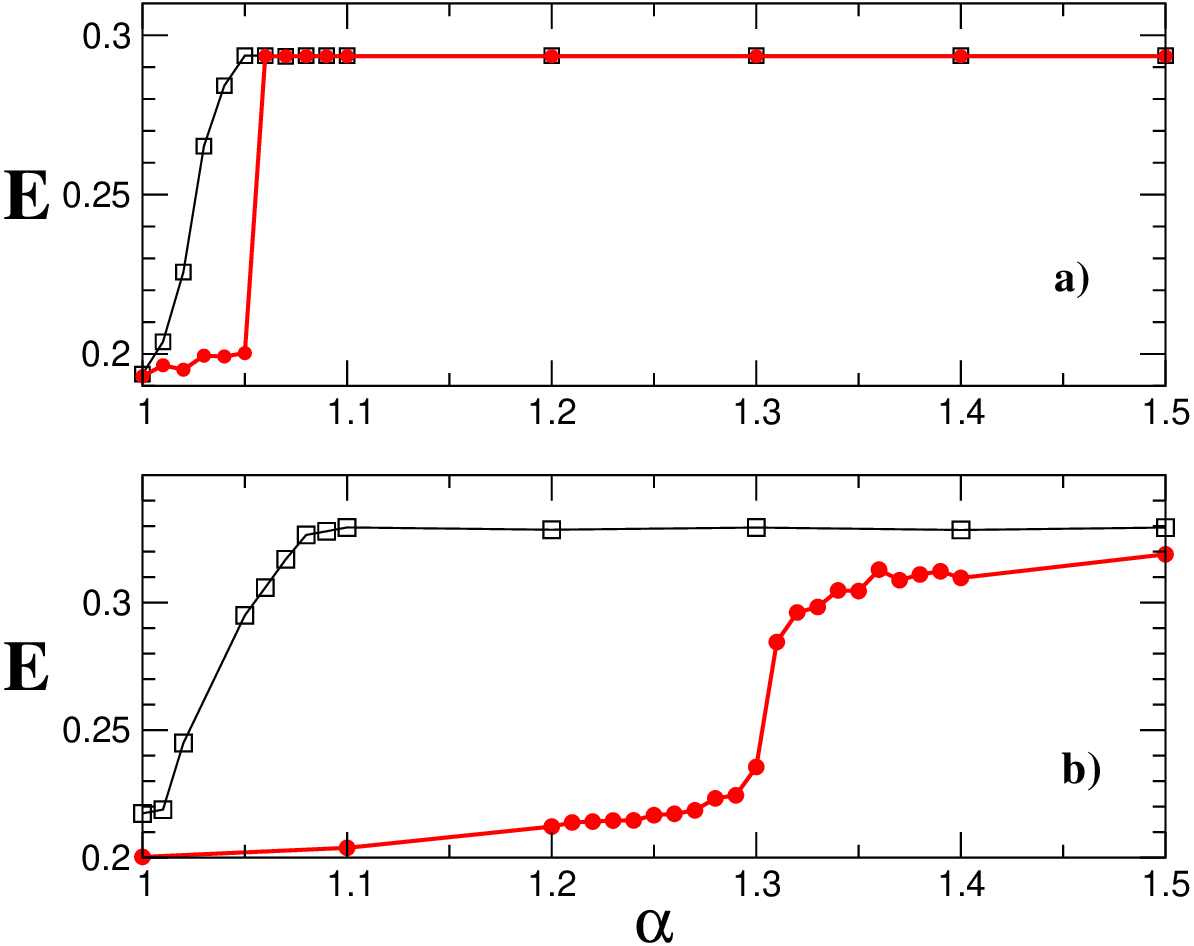}   
\caption{\label{fig2} 
Cascading failure in a) ER random graphs and b) BA scale-free networks  
as triggered by the removal of a node chosen at random 
(squares), or by the the removal of the node with largest load 
(full circles). We report the final (after the cascade) 
efficiency $E$ of the network as a function of the 
tolerance parameter $\alpha$. Both the networks considered have 
$N=2000$ and $K=10000$. 
In the case triggered by the removal of a node chosen 
at random the curve corresponds to an average over 10 triggers. 
}
\end{figure}

\bigskip
We illustrate how our model works in practice by considering 
two artificially created network topologies: 
1) Erd\"os-R\'enyi (ER) random graphs \cite{erdos}; 
2) scale-free networks, i.e. graphs with an algebraic 
distribution of degree $P(k)\sim k^{-\gamma}$ with $\gamma=3$ 
generated according to the  Barab\'asi-Albert (BA) model 
\cite{ba_model}.  
\\
In both cases we have constructed networks with $N=2000$ and $K=10000$.  
In fig.\ref{fig1} we report the typical time evolution of 
the network efficiency for the BA scale-free network.   
The dynamics of redistribution of flows is  
triggered by the removal at time $t=0$ of a node chosen at random.  
We show the results for three values of the tolerance parameter, namely 
$\alpha=1.3, 1.05, 1.01$. 
In the first case the efficiency of the network is completely 
unaffected by the failure of the node. In the second case the network 
reaches a stationary state with an efficiency lower than the initial one.  
In the third case, because of the 
lower tolerance parameter, the cascading failures collapse the system: the 
network has lost the $40\%$ of the initial efficiency. 
\\
In fig.\ref{fig2} we report the final value of 
the efficiency, i.e.\ the efficiency  after the system 
has relaxed to a stationary state, as a funtion of the 
tolerance parameter $\alpha$. 
We consider both the ER random graph and the BA scale-free graph. 
Moreover we adopt two different triggering strategies: 
$\it random ~ removals$ and $\it load-based ~ removals$. 
In the first case (squares) the node removed initially 
is chosen at random: in this way we simulate the breakdown 
of the average node of the network.  
In the second case (full circles) the removed node is a very 
special one because it is the one with the largest load. 
Both for the random and for the scale-free network we observe 
a decrease of the efficiency for small values of the tolerance 
parameter $\alpha$, and the collapse of the system 
for values smaller than a critical value $\alpha_c$. 
%We have checked that the results are qualitatively the same 
%for different realizations of the network.  
ER random graphs appear to be more resistant to cascading 
failures than BA scale-free graphs (as also found in the model 
of Ref.\cite{motter_cascade}). In both cases  
the collapse transition is always sharper for load-based removals 
than for random removals, although the values  
of $\alpha_c$ can fluctuate for different realizations. 
For the ER random graphs considered we have obtained   
$\alpha_c= 1.02 \pm 0.002$ for random removals, and 
$\alpha_c= 1.06 \pm 0.005$ for load-based removals.
For BA scale-free graphs 
$\alpha_c= 1.1 \pm 0.004$ for random removals, and 
$\alpha_c= 1.3 \pm 0.05$ for load-based removals \cite{finitesize}. 
The heterogeneity of the network plays an important role in 
the network stability. ER random graphs have an exponential load 
distribution while BA networks exhibit a power law distribution 
in the node load \cite{goh}. 
This makes a large difference between random 
removals and load-based removals in BA scale-free networks.  
In fact there are few nodes, the ones with extremely high  
initial load, that are far more likely than the other  
nodes (the most part of the nodes of network)  
to trigger cascades. 
Fig.\ref{fig2}b shows the existence of a large region 
in the tolerance parameter, $1.1 \le \alpha \le 1.3$, 
where scale-free networks are stable with respect to random removals 
and are unstable with respect to load-based removals. 
If, for instance the nodes work with a tolerance of 
$30\%$ above the standard load ($\alpha = 1.3$), the 
network is in general very stable to an initial shock consisting 
in the breakdown of a node. This means that in most of the cases 
the failure is perfectly tolerated and reabsorbed by the system.  
However, there is always a finite, although very small 
%(in this case about 0.002) 
probability that the failure triggers an avalanche 
mechanism collapsing the whole network. 
\begin{figure}[!h]
\includegraphics[width=7.cm,angle=0]{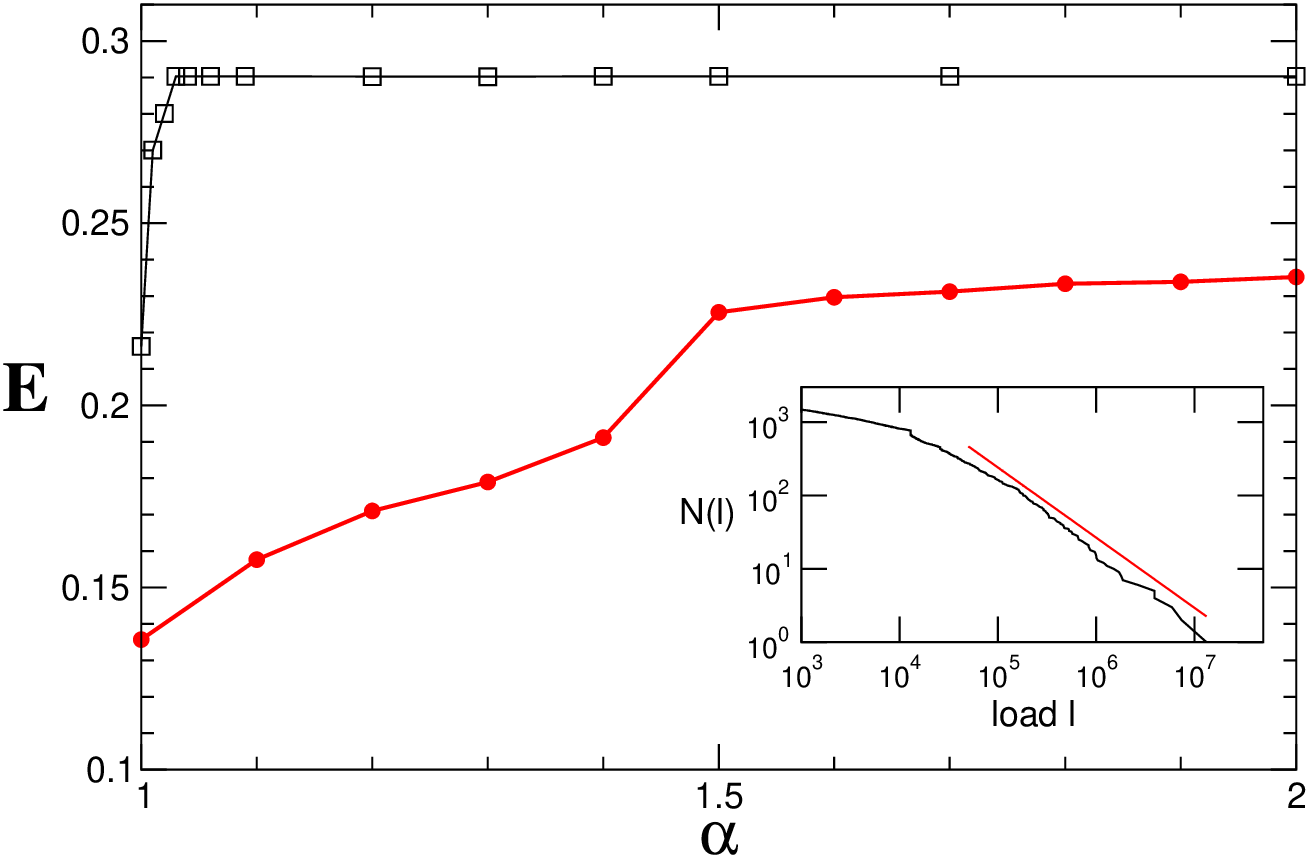}   
\caption{\label{fig3} 
Cascading failure in the Internet. The network considered is taken 
from ref.\cite{moat} 
%and has $N=6474$ nodes and $K=12567$ arcs. 
%%$<k>=3.88$ 
For each value of $\alpha$ we report the efficiency $E$ 
after the cascade triggered by the removal of a node chosen 
at random (squares), or by the removal of the node with largest load 
(full circles). 
The curve reported for random removals is an average 
over 10 different nodes. 
In the inset we plot the cumulative node load distribution.  
}
\end{figure}

\begin{figure}[!h]
\includegraphics[width=7.cm,angle=0]{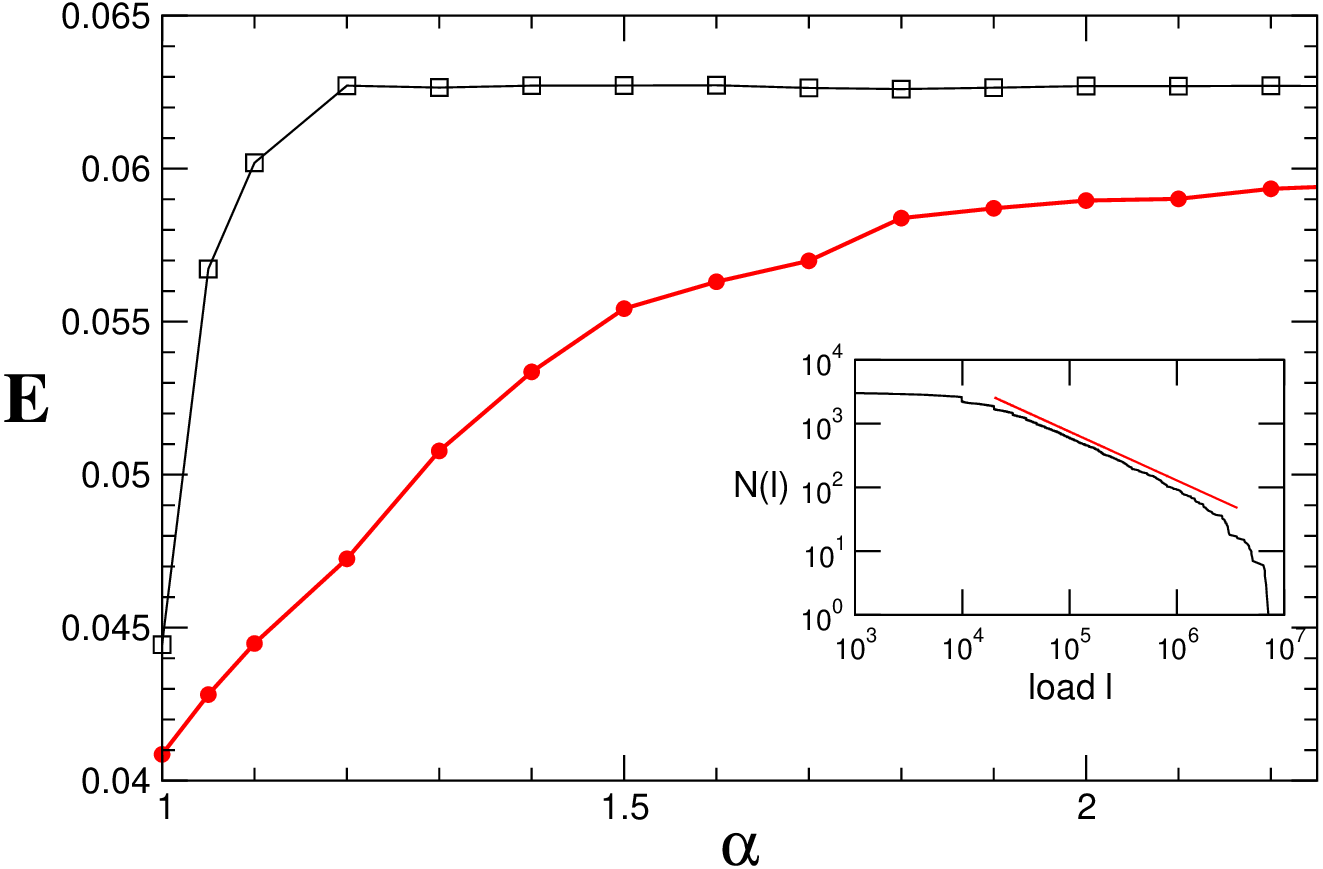}   
\caption{\label{fig4} 
Cascading failure in the electrical power grid of the 
western United States from ref.\cite{wattsandstrogatz}. 
%The network has $N=4941$ and $K=6592$. 
Same plot as in fig.\ref{fig3}
%$<k>=2.67$
}
\end{figure}

As examples from the real world we study a network of the Internet 
(at the autonomous system level \cite{mendesbook,vespignani})  
with $N=6474$ nodes and $K=12567$ arcs taken from ref.\cite{moat},  
and the electrical power grid of the 
western United States from ref.\cite{wattsandstrogatz} having 
$N=4941$ and $K=6592$. 
Although the Internet exhibit a power law degree distribution 
(as for BA scale-free networks) 
while the electrical power grid has an exponential degree 
distribution (as for ER random graphs),  
we have checked that both the networks considered are very 
hetereogeneuos from the point of view of the loads on nodes. 
In the insets of fig.\ref{fig3} and fig.\ref{fig4} we 
report $N(l)$, the number of nodes with a load larger than $l$,  
as a function of $l$: the straight lines indicate that 
the load distribution is 
consistent with a power-law with exponents respectively of 
$1.80$ and $1.75$. 
In the same figures we report the value of the 
efficiency after the cascade triggered by random failures and 
load-based failures. 
Due to the presence of few nodes with an extremely high initial 
load, the figures show a large range of $\alpha$ where 
the network is stable against random failures and is 
vulnerable with respect to the breakdown of the most loaded 
nodes. 
Although the latter events have a very low probability their 
occurrence may collapse the entire systems with a 
large effect on our life.  
These results are a possible explanation of the 
mechanism producing the experimentally observed 
Internet congestion collapses and the power blackouts.  
A small initial shock, as the breakdown of an Internet 
router or of an electrical substation or line may trigger 
avalanches mechanisms affecting a considerable fraction 
of a network that for years was proven to be stable 
with respect to similar shocks. 
As an example, if the electric power grid of 
the western United States of fig.\ref{fig4} 
works with a tolerance $\alpha=1.1$ ($\alpha=1.5$), a case in which the 
system is stable with respect to the failure of most of 
its nodes, the removal of a 
special node, the one with highest initial load, 
produces a drop of the $30\% (15\%)$ of its efficiency.

\bigskip
Summing up, in this paper we have introduced a simple model 
to explain why large but rare cascade triggered by small initial 
shocks are present in most of the 
complex communication/transportation networks  
that are the basic components of our lives. 
The model is based on a dynamical redistribution of the flow 
triggered by the initial breakdown of a component of the system.  
The results show that the breakdown of a single node
is sufficient to affect the efficiency of a network up to the 
collapse of the entire system if the node is among the ones with 
largest load. 
This is particularly important for networks with a highly 
hetereogeneous distribution of node loads as BA scale-free networks, 
but also real-world networks as the Internet 
and electrical power grids. 
Our results show that it is only the breakdown of a selected 
minority of the nodes that can trigger the collapse of the 
system. It is so the same fact that for the majority of the 
nodes nothing harmful happens, that leads us to the erroneous 
believe that our communication/transportation networks are safe. 
Therefore it should be advisable to take into proper account,   
in the design of any complex network, the 
cascading failures effects analyzed here.

{\bf Acknowledgement}. 
We thank D.J. Watts for the US power-grid data 
from ref.\cite{wattsandstrogatz} and A. Rapisarda for useful 
comments.


\begin{thebibliography}{99}
\bibitem{note0} 
The model can be easily generalized to directed graphs.
\bibitem{note1}
In the case of an unweighted graph $e_{ij}$ is $1$ if there is 
an arc joining node $i$ to node $j$, and $0$ otherwise.
\bibitem{hc} 
The harmonic composition of $N$ numbers $x_1,x_2,...,x_N$  
is defined as $( \sum_{i=1}^N \frac{1}{x_N} )^{-1}$
\bibitem{note2} 
Another possibility which gives the same results is to set 
$C_i = \alpha ~\cdot \max_j L_j(0) ~~\forall i=1,2,...,N$.
\bibitem{note3} 
As a initial damage on the network here we consider the removal of 
a node, altough the dynamics of the model can be started 
by the removal of one arc if one wants to simulate, 
for instance, the breakdown of a line in an electrical  
power grid.
\end{thebibliography}
\end{document}